\documentclass[useAMS,usenatbib]{mn2e}
\usepackage{graphicx}
\usepackage{aas_macros} 
\usepackage{times}
\usepackage[fleqn]{amsmath}
\usepackage[T1]{fontenc}  
\usepackage{aecompl}

\newcommand\msun{\rm{M_{\odot}}}

\def\stacksymbols #1#2#3#4{\def\theguybelow{#2}
        \def\verticalposition{\lower#3pt}
        \def\spacingwithinsymbol{\baselineskip0pt\lineskip#4pt}
        \mathrel{\mathpalette\intermediary#1}}
\def\intermediary #1#2{\verticalposition\vbox{\spacingwithinsymbol
        \everycr={}\tabskip0pt
        \halign{$\mathsurround0pt#1\hfil##\hfil$\crcr#2\crcr
                \theguybelow\crcr}}}

\def\lta{\stacksymbols{<}{\sim}{2.5}{.2}}
\def\gta{\stacksymbols{>}{\sim}{2.5}{.2}}

\title[Quenching the soft X-ray spectrum of clusters]
{Shaping the X-ray spectrum of galaxy clusters with AGN feedback and turbulence}

\author[M.~Gaspari]{M. Gaspari$^{1}$\\
$^{1}$Max Planck Institute for Astrophysics, Karl-Schwarzschild-Strasse 1, D-85741 Garching, Germany; e-mail: mgaspari@mpa-garching.mpg.de}

\begin{document}

\pagerange{\pageref{firstpage}--\pageref{lastpage}} \pubyear{2015}
\maketitle
\label{firstpage}

\begin{abstract}
The hot plasma filling galaxy clusters emits copious X-ray radiation.
The classic unheated and unperturbed cooling flow model predicts dramatic cooling rates and an isobaric X-ray spectrum with constant differential luminosity distribution. 
The observed cores of clusters (and groups) show instead a strong deficit of soft X-ray emission:
$dL_{\rm x}/dT \propto (T/T_{\rm hot})^{\alpha=2\pm1}$. Using 3D hydrodynamic simulations, we show that such deficit arises from the tight self-regulation between thermal instability condensation and AGN outflow injection:
condensing clouds boost the AGN outflows, which quench cooling as they thermalize through the core.
The resultant average distribution slope is $\alpha$\,$\simeq$\,2, oscillating within the observed 1\,$<\alpha<$\,3. 
In the absence of thermal instability,
the X-ray spectrum remains isothermal ($\alpha \gta 8$), while unopposed cooling drives a too shallow slope, $\alpha<1$. 
AGN outflows deposit their energy inside-out, releasing more heat in the inner cooler phase;
radially distributed heating alone induces a declining spectrum, $1<\,$$\alpha\,$$<2$. 
Turbulence further steepens the spectrum and increases the scatter:
the turbulent Mach number in the hot phase is subsonic, while it becomes transonic in the cooler phase, making perturbations to depart from the isobaric mode. Such increase in $d\ln P/d\ln T$ leads 
to $\alpha\approx3$.
Self-regulated AGN outflow feedback can address 
the soft X-ray problem through the interplay of heating and turbulence.
\end{abstract}

\begin{keywords}
hydrodynamics -- instabilities -- turbulence -- galaxies: active -- galaxies: clusters: intracluster medium -- galaxies: jets.
\vspace{-0.5cm}
\end{keywords}

\section{Introduction}\label{s:intro}
Unbalanced radiative cooling of plasma 
would lead to a catastrophic cooling flow (CF) in the core of massive 
galaxies, groups, and clusters ($r<0.1\,R_{500}$). 
In galaxy clusters, the rate of the cooling gas 
would reach unrealistic values up to $\dot M\sim10^3$ $\msun$ yr$^{-1}$ (\citealt{Fabian:1994,Peres:1998}).
The central cooling gas would monolithically condense out of the hot phase, leading to a dramatic increase in the core density and to a steady inflow due to the loss of pressure support against gravity.
In the standard CF model,
the CF is assumed to develop in isobaric mode.
The mass deposition (or inflow) rate is thus independent of temperature (\S\ref{s:res})
and associated with a constant differential luminosity distribution
\begin{equation}\label{e:isob_intro}
\frac{dL}{dT}=\frac{5}{2}\frac{\dot Mk_{\rm b}}{\mu m_{\rm p}}, 
\end{equation}
with atomic weight $\mu\simeq0.62$ for the intracluster medium (ICM).

In the last two decades, numerous observations in X-ray, UV, optical, and radio band have ruled out the presence of
standard CFs in observed systems (\citealt{Peterson:2006}; \citealt{Bohringer:2010}). 
The major CF issues are twofold:
the mass sink problem and the soft X-ray problem. 
The former is the most studied problem, arising from the fact that we do not observe very large amount of warm (H$\alpha$, [NII]; \citealt{McDonald:2011a, Werner:2013}) and cold molecular gas (CO, H$_2$; \citealt{Edge:2001,Salome:2003}) dropped out of the hot phase. 
A pure CF of almost $10^3$ $\msun$ yr$^{-1}$ would indeed double the mass of the central galaxy in less than 1 Gyr, quickly reaching $M_{\rm cold}\sim10^{12}\ \msun$ 
(\citealt{Silk:1976,Cowie:1977_CF,
Mathews:1978,Ciotti:1991,Fabian:1994} for a comprehensive review on classic CF). Tightly related to the mass sink problem, the star formation rates appear to be quenched within a few per cent of the pure cooling rate (\citealt{Odea:2008}).

The previous observations indicate that the pure cooling rates must be suppressed by one to two orders of magnitude.
Early works in the 1990s (e.g. \citealt{Binney:1995,Ciotti:1997,Tucker:1997,Silk:1998})
suggested that the required heating comes from the active galactic nucleus (AGN).
With the advent of {\it XMM-Newton} and {\it Chandra}, it became clear that the AGN feedback -- with powers up to $10^{46}$ erg s$^{-1}$ --
plays the fundamental role in the thermodynamical evolution of gaseous atmospheres
(\citealt{McNamara:2007,McNamara:2012}).
In the last years, we tested different self-regulated AGN feedback models and compared them with data (\citealt{Gaspari:2009,Gaspari:2011a,Gaspari:2011b,Gaspari:2012a,Gaspari:2012b}). The best consistent model is mechanical AGN feedback, in the form of massive subrelativistic outflows and self-regulated by the accretion of cold gas on to the black hole (\citealt{Gaspari:2013_rev,Gaspari:2014_scalings}).
The cooling rates are quenched by 20 fold, at the same time preserving the cool-core structure
(as the positive temperature gradient).
The rapid intermittency and high accretion rates of cold accretion permit the recurrent generation of buoyant bubbles, 
weak shocks, and the uplift of metal-rich low-entropy gas. The driven AGN turbulence is crucial for the growth of nonlinear thermal instability (TI): the condensation of extended filaments boosts the accretion rate, closing the self-regulated feedback loop (known as chaotic cold accretion mechanism -- CCA; \citealt{Gaspari:2013_cca,Gaspari:2014_rot}).
Such results have been independently corroborated by other observational and numerical investigations 
(\citealt{Sharma:2012,Li:2014,Voit:2015a,Voit:2015b}). 

While the mass sink problem has been extensively tackled,
the problem related to the deficit of emission in the
soft X-ray band remains less studied. The high spectral resolution of {\it XMM} 
Reflection Grating Spectrometer (RGS) has revealed that the residual CF is
not constant as a function of temperature. 
The emission measure of the X-ray spectrum shows a substantial deficit
below 2 keV, typically correlated with the presence of multiphase structures 
(\citealt{Kaastra:2001,Peterson:2001,Tamura:2001,Sakelliou:2002}).
Combining the observations of the last decade (see \S\ref{s:res}),
the X-ray spectrum written as $dL_{\rm x}/dT\propto (T/T_{\rm hot})^\alpha$ 
is constrained to have a slope $\alpha\approx2\pm1$
(\citealt{Peterson:2003,dePlaa:2004,Kaastra:2004,Sanders:2009,Grange:2010,Sanders:2010}).

In this work, we probe the impact of self-regulated AGN feedback
in shaping the X-ray spectrum
through the combined action of heating and turbulence.
Direct heating reduces the average cooling rate.
However, how can it deposit more energy in the cooler and denser gas 
with increasing emissivity?
Does turbulence play a role in the cold phase thermodynamics?
We propose that isobaric radiative cooling is inapplicable in a turbulent ICM atmosphere, even
if motions are subsonic (\S\ref{s:res}). Both simulations and observations have shown that the ICM is turbulent 
(\citealt{Schuecker:2004,Nagai:2007,Vazza:2009,Gaspari:2013_coma,Gaspari:2014_coma2}).
Turbulence not only plays central role in the formation of nonlinear TI and dense filaments,
but also makes the cooler phase to depart from the isobaric mode.
As motions become transonic in the cold phase, this effect becomes increasingly more important at lower temperatures, contributing to the observed decline in the soft X-ray spectrum.

\vspace{-0.5cm}

\section[]{Physics \& Numerics} \label{s:init}
The setup and equations of the 3D hydrodynamic simulations with self-regulated AGN feedback are
described in \citeauthor{Gaspari:2012a} (\citeyear{Gaspari:2012a}; hereafter G12).
Here, we summarize the essential physics and numerics.

We adopt as reference system a massive galaxy cluster with virial mass $\approx$\,$10^{15}\ \msun$ and baryon fraction $0.15$. The system is modelled on the well-studied cluster A1795 by using the observed cool-core temperature profile (\citealt{Ettori:2002}). The initial minimum core temperature is $\simeq$\,3 keV; the self-regulated AGN feedback preserves it at $\sim$\,3\,-\,4 keV also during the evolution
(the background profile related to the diffuse phase is thus not affecting the soft X-ray decline).
The cluster is initialized in hydrostatic equilibrium in the Navarro-Frenk-White (NFW)
potential dominated by dark matter. The initial profiles are perturbed (due to cosmological flows) 
to avoid idealized symmetric conditions, albeit the fluctuations are soon driven by the AGN turbulence.
We use FLASH4 code with third-order accurate piecewise parabolic method (PPM) to integrate the hydrodynamic equations
(equations 1-4 in G12).
The domain is 1.27 Mpc on a side, covered by static mesh refinement zooming in the cluster core. The maximum resolution is 300 pc. 
The fuelling of the black hole is dominated by cold accretion (\S1).
Self-consistent accretion simulations down to 20 gravitational radii
showed that in CCA most of the cold gas condensing within 10 kpc is accreted in a few free-fall times (\citealt{Gaspari:2013_cca,Gaspari:2014_rot}), irrespectively of the Bondi radius, which is only meaningful for hot/Bondi accretion models.
We ran a shorter simulation with double resolution and found very similar results, in particular in terms of the $dL_{\rm x}/dT$ diagnostic. The X-ray range ($\gta5\times10^6$ K) describing $dL_{\rm x}/dT$ (Fig.~\ref{fig:xs_pjet}) is highly resolved. The smallest $10^4$ K clouds approaching resolution may be diffused, but these have no impact on $dL_{\rm x}/dT$ and are negligible for the total accretion rate.

In addition to hydrodynamics, we include two key physical source terms: radiative cooling and self-regulated AGN outflows. The ICM emits radiation via bremsstrahlung ($T>1$ keV) and line cooling ($T<1$ keV), with total emissivity $\mathcal{L}=n_{\rm e}n_{\rm i}\Lambda(T,Z)$, where $n_{\rm e}$ and $n_{\rm i}$ are the electron and ion number density, respectively ($\mu_{\rm e}\mu_{\rm i}\simeq1.464$). The cooling function $\Lambda(T,Z)$ is based on the tabulated values by \citet{Sutherland:1993} for a fully ionized plasma with metallicity $Z=0.3$ Z$_\odot$.
The stable cold phase is preserved at $10^4$ K. The split source term is integrated with an exact method, which tracks the cooling gas with high accuracy.

Unopposed radiative cooling would lead to a massive CF catastrophe.
As shown by observations and simulations (\S1),
mechanical AGN feedback is the preferred mechanism to quench CFs (and subsequent star formation). 
Bipolar AGN outflows 
are injected
through the internal boundaries in the middle of the domain
with kinetic power
\begin{equation}\label{e:Pjet}
P_{\rm out} = \epsilon\,\dot M_{\rm acc}\,c^2,
\end{equation}
where $\epsilon$ is the kinetic efficiency and $\dot M_{\rm acc}$ is the black hole accretion rate
tied to the sinked gas in the inner zones. 
As shown in \citet{Gaspari:2011a,Gaspari:2014_scalings}, and G12, 
the long-lived presence of observed cool cores and the absence of severe overheating (negative $T(r)$ gradients)
put tight constraints on the consistent kinetic efficiency, with best value $\epsilon\simeq5\times10^{-3}$.
Besides simulations (see also \citealt{diMatteo:2005}), observational works studying the AGN kinetic luminosity function
also retrieve $\epsilon\simeq5\times10^{-3}$ (cf.~\citealt{Merloni:2008,LaFranca:2010}).
Most of the accreted gas is in the form of cold clouds and filaments, while the hot phase accounts for just a few per cent of the accretion rate 
(CCA; \citealt{Gaspari:2013_cca}). The typical mass rate and velocity of the outflows are a few $\msun$ yr$^{-1}$ and $5\times10^4$ km s$^{-1}$, respectively.

To disentangle the dominant physics, we perform two additional runs: one with purely distributed heating and one
with distributed heating plus forced turbulence. 
Computationally, we set the heating rate (per unit volume) to be equal to the average radiative emissivity in finite radial shells, $\mathcal{H}\approx\langle\mathcal{L}\rangle$. Forced turbulence is driven via a spectral forcing scheme based on an Ornstein-Uhlenbeck random process, which drives a time-correlated and zero-mean acceleration field. The driven velocity dispersion is $\sigma_v\sim250$ km s$^{-1}$ (with injection scale $L\approx80$ kpc).
Both modules are described in detail in \citealt{Gaspari:2013_cca} (section 2).

\vspace{-0.5cm}

\section[]{Results \& Discussion} \label{s:res}
\begin{figure} 
       \includegraphics*[scale=0.45]{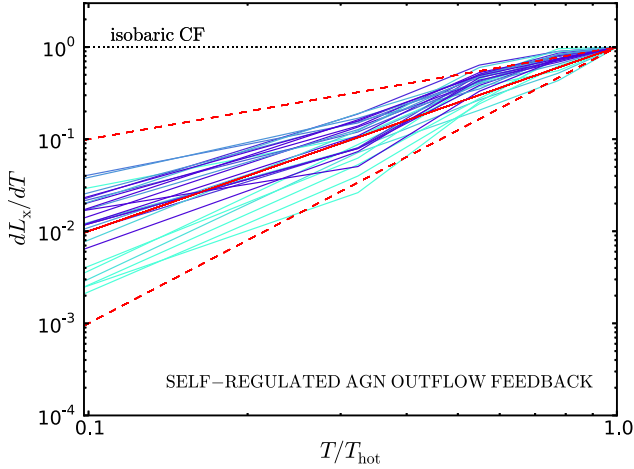}
     \caption{The X-ray spectrum in the form of differential luminosity distribution per temperature interval, for the reference self-regulated AGN outflow feedback model ($\epsilon\simeq5\times10^{-3}$). As in observations (\S1), we use the hot gas temperature in the core for the normalization ($T_{\rm hot}\simeq 4$ keV) and we extract the distribution
     within the cooling radius, $r<100$ kpc ($\sim$1 arcmin for A1795). The distribution is normalized to the pure isobaric CF (dotted; equation \ref{e:isob}).
     The colour coding is related to the accretion rates averaged over 20 Myr after each snapshot time (cyan to blue $\approx$\,1\,-\,20 $\msun$ yr$^{-1}$). The simulated spectrum is consistent with the observational constraints (red; $\alpha=2\pm1$, equation \ref{e:obs}), showing a strong deficit down to $T\simeq0.4$ keV (we cut the plot to this value as it roughly corresponds to {\it Chandra} lower sensitivity).     
      The departure from the standard isobaric CF naturally develops via the tight self-regulation between TI condensation and turbulent AGN heating.}
          \label{fig:xs_pjet}
\end{figure}

In Fig.~\ref{fig:xs_pjet}, we present the key diagnostic plot defining the X-ray spectrum: 
the differential luminosity distribution per temperature interval. The $dL_{\rm x}/dT$ distribution contains crucial information.
The energy equation (or first law of thermodynamics) affected by radiative losses and heating
can be written in Lagrangian form as
\begin{equation}\label{e:en}
\rho\,\frac{d\varepsilon}{dt} - \frac{P}{\rho}\frac{d\rho}{dt} = -\,n_{\rm e}n_{\rm i}\, \Lambda + n\,\Gamma, 
\end{equation}
where $\varepsilon=k_{\rm b}T/[\mu m_{\rm p}(\gamma-1)]$ is the gas specific thermal energy, $P=k_{\rm b} \rho T/\mu m_{\rm p}$ is the pressure, $\gamma=5/3$ is the adiabatic index,
and $\Gamma\equiv d E/d t$ is the injected heating rate per particle (erg~s$^{-1}$).
We focus in this study on the differential X-ray luminosity, which is defined as
$dL_{\rm x}\equiv\,n_{\rm e}n_{\rm i}\, \Lambda(T,Z)\,dV$ (\citealt{Peterson:2006}), where $\Lambda(T,Z)$ is the cooling function (see \S2).
Using the classic definition $\dot M\equiv-\,\rho\,dV/dt$ in equation (\ref{e:en}), yields
\begin{equation}\label{e:en2}
\frac{dL_{\rm x}}{dT}=\dot M \left(\frac{\gamma}{\gamma-1}\frac{k_{\rm b}}{\mu m_{\rm p}}-\frac{1}{\rho} \frac{dP}{dT} -\frac{1}{\mu m_{\rm p}}\frac{dE}{dT}\right)
\end{equation}
\begin{equation}\label{e:dLdT}
\ \ \ \ \ \ \ \ \ =\frac{\dot M k_{\rm b}}{\mu m_{\rm p}} \left(\frac{5}{2}- \gamma_T - \frac{1}{k_{\rm b}}\frac{dE}{dT}\right),
\end{equation}
where $\gamma_T\equiv d\ln P/d\ln T=\gamma_{\rm eff}/(\gamma_{\rm eff}-1)$; 
$\gamma_{\rm eff} = d\ln P/d\ln \rho$ is known as the effective adiabatic index.
Equation (\ref{e:dLdT}) indicates $dL_{\rm x}/dT$ can be reduced
by changing the effective adiabatic index (e.g. turbulence) and/or an external injection of heat (last term).

Let us neglect the heating term for the moment ($dE=0$).
If cooling proceeds in isobaric mode (as in the standard CF), then $\gamma_T=0$ ($\gamma_{\rm eff} = 0$) and
the luminosity distribution in equation (\ref{e:dLdT}) depends only on a given mass inflow rate 
(dotted horizontal line in Fig.~\ref{fig:xs_pjet}):
\begin{equation}\label{e:isob}
\frac{dL_{\rm x}}{dT}=\frac{5}{2}\frac{\dot Mk_{\rm b}}{\mu m_{\rm p}} \quad \quad \quad \quad \quad \quad  \quad \quad \quad \quad{\rm [isobaric].}
\end{equation}
Note that the X-ray spectrum can be interpreted in terms of differential emission measure
$dY/dT= \Lambda^{-1}\, dL_{\rm x}/dT$.
In the isobaric CF, $dY/dT\propto T$, since $\Lambda\propto T^{-1}$
and $dL/dT \propto \dot M$ is constant.
If cooling develops instead in isochoric mode (constant density), $\gamma_T=1$ ($|\gamma_{\rm eff}| = \infty$),
the luminosity distribution is reduced to 
\begin{equation}\label{e:isoc}
\frac{dL_{\rm x}}{dT}=\frac{3}{2}\frac{\dot Mk_{\rm b}}{\mu m_{\rm p}} \quad \quad \quad \quad \quad \quad  \quad \quad \quad \quad{\rm [isochoric].}
\end{equation}
In the extreme adiabatic mode, $\gamma_T=5/2$ ($\gamma_{\rm eff} = 5/3$),
the luminosity distribution can be strongly suppressed:
\begin{equation}\label{e:adia}
\frac{dL_{\rm x}}{dT} \ll \frac{\dot Mk_{\rm b}}{\mu m_{\rm p}}   \ \,\, \quad \quad \quad \quad \quad \quad  \quad \quad \quad \quad{\rm [adiabatic].}
\end{equation} 

\begin{figure} 
       \includegraphics*[scale=0.45]{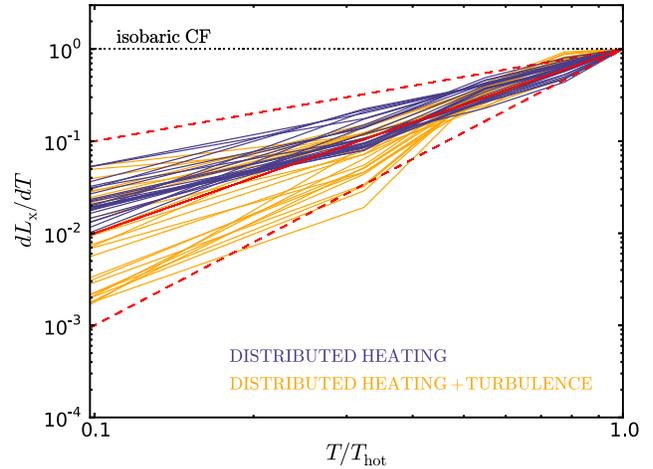}
     \caption{Differential X-ray luminosity distribution for the run with distributed heating (no turbulence; dark blue)
     and the same model with the addition of forced turbulence ($\sigma_v\,$$\sim$$\,250$ km s$^{-1}$; orange). Analogue of Fig.~\ref{fig:xs_pjet}. The average cooling rate is comparable to the outflow feedback models, $\approx30\ \msun$ yr$^{-1}$. 
     Direct heating is fundamental to produce a substantial decline in the soft X-ray spectrum.
     Since the AGN heating distribution is a decreasing function of radius, it is naturally tied to the cooler, inner gas, leading to a positive slope $\alpha$. Turbulence increases the importance of adiabatic processes and mixing in the cooler phase (where it becomes transonic), further widening the departure from the isobaric regime     
     ($\gamma_T>0$; equation \ref{e:dLdT}). 
     The $\gamma_T$ increase steepens on average the spectrum slope, with values up to $\alpha\approx 3$, while the scatter is now raised to $\sim\,$1 dex at low temperatures.}
          \label{fig:xs_gTE}
\end{figure}

Which are the constraints provided by observations?
\citet{Kaastra:2004} studied $dY/dT$ in the core of 17 well-observed CF clusters. The average sample slope of the emission measure distribution is $\alpha' \approx 3.1$,
which translates to a luminosity slope $\alpha=\alpha'-1\approx2.1$. The scatter is significant, with most clusters contained within $1<\,$$\alpha\,$$<3$. A few objects have extremely low and high values, such as Perseus ($\alpha\approx0.5$), A1835 ($\alpha\sim0$), MKW9 ($\alpha\approx8$), and A399 ($\alpha\rightarrow\infty$, isothermal).
\citet{Peterson:2003} also rule out the standard CF, finding multitemperature spectra with $\alpha\approx1$\,-\,2 in 14 clusters (fig.~9).
\citet{dePlaa:2004} measure $\alpha\simeq1.63$ in the core of A478. \citet{Grange:2010} retrieve $\alpha\approx1.1$ in J1539 cluster. \citet{Sanders:2009} find a soft X-ray deficit consistent with
$\alpha\sim1.5$ in 2A 0335+096. 
The X-ray bright cool-core clusters A262, A3581, and HCG62 show a multiphase structure with $\alpha\sim2$ (\citealt{Sanders:2010}), although the distribution is more complex than a simple power law. In Centaurus cluster, the RGS spectrum has $\alpha\simeq1.31$ (\citealt{Sanders:2008_Cen}),
while M87 spectrum is fitted by $\alpha\simeq1.13$ (\citealt{Werner:2006}).
Massive groups similarly display suppressed soft X-ray emission. For instance, in NGC 5044,
\citet{Tamura:2003} retrieve $\alpha\sim3$ below 1 keV, flattening near 0.3 keV.
To summarize, the observed deficit in the soft X-ray luminosity distribution can be written as
\begin{equation}\label{e:obs}
\frac{dL_{\rm x}}{dT}\simeq\frac{5}{2}\frac{\dot Mk_{\rm b}}{\mu m_{\rm p}} \left(\frac{T}{T_{\rm hot}}\right)^\alpha,
\end{equation} 
where $\alpha=2\pm1$ (red lines in Figs \ref{fig:xs_pjet} and \ref{fig:xs_gTE}) and $T_{\rm hot}$ is the ambient hot temperature in each cluster core.

In Fig.~\ref{fig:xs_pjet} (blue and cyan), we show the differential X-ray luminosity distribution resulting from the reference AGN feedback simulation during tight self-regulation (\S\ref{s:init}), extracted in the cluster core ($r<100$ kpc) and sampled every 50 Myr.
The differential luminosity decreases from the core ambient temperature 
($T_{\rm hot}\simeq4$ keV) to the soft X-ray floor ($\simeq$\,0.4 keV), reaching a deficit of 2 dex compared with
the pure isobaric CF (dotted black). 
The simulated $dL_{\rm x}/dT$ slope oscillates around $\alpha\simeq 2$, with a scatter contained within 1 and 3, which is
consistent with the observed values (equation \ref{e:obs}). 
The steady oscillation around $\alpha\approx2$ is reached when the system is
tightly self-regulated, dancing between periods of TI condensation and outflow injection.
Perturbations in the heated atmosphere grow quickly nonlinear as the `TI-ratio' $t_{\rm cool}/t_{\rm ff}<10$
(G12, sec.~2), i.e. in the region $r\lta 20$ kpc for A1795.
Cold clouds and extended filaments condense out of the hot phase, rain on to the black hole, boosting
the accretion rate and the AGN outflow power. Correlating $dL_{\rm x}/dT$ with the accretion rate averaged over 20 Myr after each snapshot time, we find that higher values are linked to shallower $dL_{\rm x}/dT$. 
Since the evolution is self-regulated, we are witnessing the continuous fuelling of the black hole via cold accretion. The stronger the TI condensation (lower $\alpha$), the larger the amount of fuel, subsequently boosting $P_{\rm out}$
(up to $\sim\,$$10^{45}$ erg s$^{-1}$).

We analysed other runs departing from proper self-regulation.
In the absence of TI
(e.g. in runs with $\epsilon > 0.01$ or $t_{\rm cool}/t_{\rm ff} \gg 20$), the X-ray spectrum approaches the 
isothermal behaviour with $\alpha \approx 8$ (not shown).
In the other extreme run with pure cooling, we find a shallow spectrum slope, $\alpha \approx 0.7$. We notice that even without heating, a cooling halo develops nonlinear TI in the presence of relevant perturbations (e.g. if the amplitude is $> 0.1$; \citealt{Gaspari:2013_cca,Gaspari:2014_rot}). 
The TI condensation process and the preexisting perturbations drive chaotic motions ($\sigma_v\sim70\ \rm{km\, s}^{-1}$)
leading to the departure from the pure isobaric mode ($\alpha > 0$; see below).
Moreover, below 1 keV, $t_{\rm cool}$ starts to fall below the sound crossing time $t_{\rm c_{\rm s}}$, helping the perturbation mode to shift to the isochoric regime (equation \ref{e:isoc}).
A soft X-ray decline is thus 
associated with the formation of TI and a multiphase cluster core.

The irreversible heating due to the AGN feedback
is a key component 
suppressing the luminosity distribution (last term in equation \ref{e:dLdT}). 
Why should $dE/dT$ increase at lower $T$?
Simulations show that AGN feedback is an inside-out process, depositing more energy in the inner regions
and less energy at larger radii. 
The AGN outflows are injected through the inner boundary and the simulations self-consistently compute the
complex nonlinear thermalization of the outflows through the cluster core.  
As shown in G12 (fig.~9, right), such energy deposition is maximal in the central few kpc and decreases with radial distance (bottom to top panels). Below, we test forced radially distributed heating, which results in a similar $dL_{\rm x}/dT$, further corroborating that the more realistic AGN outflow feedback simulations approach the inside-out heating distribution.
Since cooling strongly increases at smaller radii
and condensed TI sink to the centre,
AGN heating naturally targets the inner cooler regions, inducing larger $dE/dT$ at lower temperatures, thereby
steepening the
differential luminosity distribution (Fig.~\ref{fig:xs_pjet}).

On top of external AGN heating, is turbulence
contributing to suppressing the soft X-ray spectrum?
In the $dE=0$ case, in order to match observations (equation \ref{e:obs}),
the $\gamma_T$ index (equation \ref{e:dLdT}) has to vary as
\begin{equation}\label{e:gT_obs} 
\gamma_T \approx \frac{5}{2}\,\left[1-\left(\frac{T}{T_{\rm hot}}\right)^\alpha\right], 
\end{equation} 
i.e. $\gamma_T$ should monotonically increase
with decreasing temperature. 
In other words, the gas dynamics should shift from the isobaric towards the isochoric/adiabatic mode at low $T$.
It has been shown that the thermodynamic mode driven by turbulence changes with the Mach number
(\citealt{Gaspari:2013_coma,Gaspari:2014_coma2}), shifting from the isobaric to adiabatic mode
as ${\rm Ma}> 0.5$. The subsonic motions driven in the hot phase become 
transonic/supersonic in the condensing cooler phase.
This increases the importance of the non-isobaric modes (see also interstellar medium studies; \citealt{Sanchez-Salcedo:2002}), together with the impact of turbulent mixing,
which can further steepen $dL_{\rm x}/dT$.

Since the self-regulated AGN outflow models include a complex nonlinear mix of heating and turbulence,
we computed two additional control runs: one purely with heating distributed in radial shells (now an assumption, not a result) and the other with the addition of forced subsonic turbulence ($\sigma_v\sim250$ km s$^{-1}$; \S\ref{s:init}).
Fig.~\ref{fig:xs_gTE} shows that direct heating (dark blue lines) is important to induce a decline of the spectrum with
$1<\,$$\alpha\,$$<2$, thereby reproducing part of the AGN feedback evolution (Fig.~\ref{fig:xs_pjet}).
However, turbulence (orange) is necessary to broaden the distribution (up to 1 dex at low $T$) and to generate steeper X-ray spectra with slopes up to $\alpha\approx3$. The increase of $\gamma_T$ at lower temperature (equation \ref{e:gT_obs}) via turbulence is thus an important component to address the soft X-ray problem.

Returning to the reference self-regulated run (Fig.~\ref{fig:xs_pjet}),
we computed the vorticity magnitude in the cluster core
for two random snapshots having
$\alpha \approx 1.5$ and $2.5$.
The only source of vorticity is the turbulence
excited by the AGN feedback (via outflows, bubbles, and shocks); thereby, the driven vorticity $\boldsymbol{\omega} \equiv \boldsymbol{\nabla}\times\boldsymbol{v}$ is a good tracer of turbulent motions. 
From the shallow to steeper curve, we find that the vorticity magnitude increases from $\omega\approx0.02$ to 0.04 Myr$^{-1}$. The latter value is analogous to the vorticity retrieved in the forced turbulence run with $\sigma_v\sim250$ km s$^{-1}$.
Therefore, a subsonic turbulent Mach number in the hot phase (4 keV) as small as ${\rm Ma}_{\rm hot}\sim0.25$ can have important consequences for the condensing cooler phase (${\rm Ma}_{\rm cold} \gta 1$), leading to departures from the isobaric mode and a constant differential luminosity, even in the pure cooling runs.
As further support for this scenario, we find in all runs a systematic drop of pressure in the cold phase down to $10^4$ K.
While the pure adiabatic mode ($\gamma_T=5/2$) is hard to achieve, 
the isochoric regime is the most frequent state of the cold phase. Turbulence thus requires the complementary 
AGN heating to induce substantially steep X-ray spectra (Fig.~\ref{fig:xs_gTE}).

\vspace{-0.5cm}

\section[]{Conclusions} \label{s:conc}
This work aims to study 
the impact of AGN feedback and turbulence in shaping the X-ray spectrum of galaxy cluster cores, in particular
addressing the observed deficit of emission towards the soft X-ray band. 
We combined extensive literature data of the last decade, showing that 
the observed X-ray differential luminosity distribution has a steep decline constrained as $dL_{\rm x}/dT \propto (T/T_{\rm hot})^{\alpha=2\pm1}$. This is a dramatic
deviation from the standard isobaric CF model having constant distribution (equation \ref{e:isob}).
The 3D hydrodynamic simulation with AGN outflow feedback ($\epsilon\simeq5\times10^{-3}$)
reproduces the observed soft X-ray spectrum decline
during tight self-regulation, $1<\,$$\alpha\,$$<3$.
The simulations show that the deficit naturally arises from the symbiotic link between 
cold gas condensation and AGN outflow injection: in a turbulent and heated atmosphere,
thermal instabilities grow quickly nonlinear, boosting the accretion rate and thus the AGN outflows;
the kinetic outflows gradually thermalize through the core, reheating the gas and thus quenching cooling,
in a self-regulated loop.

Without tight self-regulation, two scenarios are ruled out by soft X-ray observations.
If the system is strongly overheated (e.g. due to very high feedback efficiency or $t_{\rm cool}/t_{\rm ff}$), 
cooling and growing TI are negligible, thereby 
the spectrum remains nearly isothermal ($\alpha\gta8$). Conversely, with negligible AGN heating, 
a pure CF develops producing a shallow spectrum slope, $\alpha < 1$.
Even in the latter pure cooling regime, turbulent motions excited by the TI condensation and cosmological perturbations
prevent to reach the isobaric mode.
The drop in the soft X-ray band is thus associated with 
the formation of TI and multiphase cores.

To disentangle the main processes,
we preformed two additional control runs
forcing radially distributed heating and subsonic turbulence, 
which are self-consistently achieved in the more realistic AGN outflow runs.
Radially distributed heating alone produces a declining X-ray spectrum with slope $1<\,$$\alpha\,$$<2$.
The heat deposition is higher for lower $T$ gas, since the AGN outflows thermalize
as a decreasing function of radius (G12), in similar way as the forced distributed heating models. 
Therefore, the AGN energy is deposited relatively more in the inner cooler phase (TI condensation further advects the cold gas to the centre), producing a steep spectrum.
Forced subsonic turbulence ($\sigma_v\sim250$ km s$^{-1}$)
introduces a significant scatter (1 dex) 
and further steepens the spectrum to $2<\,$$\alpha\,$$<3$. 
While the subsonic Mach number in the hot phase becomes transonic in the cooler phase,
the perturbation mode departs from isobaric ($\gamma_T > 0$), suppressing $dL_{\rm x}/dT$ at progressively lower $T$
(equation \ref{e:dLdT}).
The self-regulated feedback simulation similarly shows that the slope steepens with increasing vorticity (tracer of turbulence), with maximum suppression reached with the vorticity of the forced stirring run.
This correlation and the drop of pressure in the cold phase corroborate the scenario of non-isobaric cooling.

To conclude, mechanical AGN feedback tightly self-regulated via CCA 
is able to quench the cooling rates by 20 fold while preserving the cool-core structure 
(the mass sink problem); at the same time,
such mechanism can induce a steep X-ray spectrum (the soft X-ray problem)
through the complementary action of AGN outflow heating and the driven subsonic turbulence.

\vspace{-0.5cm}

\section*{Acknowledgements}
The FLASH code was in part developed by the DOE NNSA-ASC OASCR Flash centre at the University of Chicago. 
MG is grateful for the Max Planck Fellowship. 
MG thanks the anonymous referee who helped to improve the manuscript.

\vspace{-0.5cm}

\bibliographystyle{mn2e}
\bibliography{biblio}

\label{lastpage}

\end{document}